# Construction of Two Statistical Anomaly Features for Small-Sample APT Attack Traffic Classification

Ru Zhang[1], Wenxin Sun[1], Jianyi Liu[1], Jingwen Li[1], Guan Lei[2], Han Guo[3]
[1]Beijing University of Posts and Telecommunications, Beijing 100876, China
[2]Tsinghua University, School of Electrical Engineering, Beijing 100091, China
[3]State Grid Information & Telecommunication Branch, Beijing 100761, China

Corresponding author: Jianyi Liu (liujy@bupt.edu.cn), Wenxin Sun (1586246996@qq.com).

This work was supported in part by the National Natural Science Foundation of China under Grant U1936216, and Grant U1836108.

**ABSTRACT** Advanced Persistent Threat (APT) attack, also known as directed threat attack, refers to the continuous and effective attack activities carried out by an organization on a specific object. They are covert, persistent and targeted, which are difficult to capture by traditional intrusion detection system(IDS). The traffic generated by the APT organization, which is the organization that launch the APT attack, has a high similarity, especially in the Command and Control(C2) stage. The addition of features for APT organizations can effectively improve the accuracy of traffic detection for APT attacks. This paper analyzes the DNS and TCP traffic of the APT attack, and constructs two new features, C2Load_fluct (response packet load fluctuation) and Bad_rate (bad packet rate). The analysis showed APT attacks have obvious statistical laws in these two features. This article combines two new features with common features to classify APT attack traffic. Aiming at the problem of data loss and boundary samples, we improve the Adaptive Synthetic(ADASYN) Sampling Approach and propose the PADASYN algorithm to achieve data balance. A traffic classification scheme is designed based on the AdaBoost algorithm. Experiments show that the classification accuracy of APT attack traffic is improved after adding new features to the two datasets so that 10 DNS features, 11 TCP and HTTP/HTTPS features are used to construct a Features set. On the two datasets, F1-score can reach above 0.98 and 0.94 respectively, which proves that the two new features in this paper are effective for APT traffic detection.

## I. INTRODUCTION

Since the Stuxnet attack in 2010, more and more Advanced Persistent Threat(APT) attacks have appeared, posing a great threat to network security. Compared with general attacks, APT attacks have the features of concealment, persistence, and pertinence. After a successful connection, APT attackers may be lurking in the target network for a long time to maintain monitoring [1], and may move laterally in the internal network. It gradually approaches the target host, so it is more difficult to detect than traditional malicious traffic. In[2][3] the authors have analysis pointed out that APT attacks are mostly carried out by organizations, with long-term penetration of specific targets, and most of them are aimed at stealing information. The traffic distribution within the same organization is similar.

The current deep-learning technology has been widely used in abnormal traffic classification and detection. Many abnormal traffic classification and detection methods using deep learning have been proposed. The combination of deep learning and other methods can improve the accuracy and versatility of abnormal traffic classification. Wu ZD [4] et al. proposed an intrusion detection method based on semantic coding and deep learning. Hassan et al. [5] have proposed an integrated learning model based on the combination of a random subspace (RS) learning method and a random tree (RT) to detect traffic attacks. Some studies [6] combine stream computing and deep learning to realize network attack detection. However, the deep learning method that includes feature extraction is more accurate compared with the method that does not include feature extraction models [7].

Therefore, in some studies, features extraction and use are an important part of their methods. Some studies[8] proposed a Convolutional Neural Networks (CNN) structure based on multipath residual learning to learn features of different granularities, to improve the expression ability of neural networks and the diversity of feature representation. Some studies [9] use a one-dimensional convolutional network to analyze the sequence features of the hybrid neural network, and uses deep neural network to learn the features of high-dimensional feature vectors, including general statistical features and environmental features, to conduct comprehensive analysis on network anomaly detection. And other studies [10] have proposed Multi-distributed Variational Autoencoder (MVAE). The tag information allows MVAE to force/divide network data samples into different categories in different regions in the potential feature space. Still some literature [11] use standardized UTF-8 character encoding for spatial feature learning (SFL), and can fully extract the features of real-time HTTP traffic without having to use encryption, calculation of entropy or compression. Other studies[12] have designed a deep neural network (DNN) FC-Net based on the meta-learning framework, which consists of a feature extraction network and a comparison network.

The samples used in the above methods are large enough, so deep learning algorithms can achieve good results. However, the number of APT attack samples is scarce, and it is difficult to meet the training needs of deep learning algorithms. Besides, the traffic features in the above methods cannot well represent the characteristics of APT attack traffic. Therefore, in order to improve the accuracy of APT attack traffic classification in the case of small sample, it is necessary to construct a feature sets for APT traffic classification.

In reality, it's difficult to prove whether these malicious traffic originated from APT attacks in the social engineering and malicious sample implantation stage of the APT life cycle. Therefore, this article focuses on the command and control(C2) connection phase. Starting from the obtained APT samples of an organization, this paper analyzes the traffic features of attack samples from the same organization, designs new traffic features for APT attacks, builds a multi-type Features set, and uses the PADASYN-AdaBoost algorithm to classify malicious traffic. We summarize our contributions as follows:

(1) Design two new features: C2Load_fluct and Bad_rate. We design two new features to improve the classification results of APT traffic. C2Load_fluct describes the correlation between the information carried by the domain name per unit length and the domain name. The lower the value of C2Load_fluct, the higher the possibility of using the domain name for information transmission. Bad_rate reflects the network status when APT malicious traffic and normal traffic are generated.

(2) Construct Features set. We combined the two new features designed with the features commonly used for abnormal traffic classification detection to construct DNS traffic feature set and TCP\HTTP/HTTPS traffic feature set. These two feature sets help to classify traffic generated by establishing connections with the remote C2 server and leaking data.

(3) Propose an novel model for APT traffic classification. We use the improved PADASYN method to balance the sample data, and optimize the AdaBoost algorithm to build a three-classifier. Use the traffic data generated by real APT samples to test our method, and the results show that the features proposed in this article can improve the accuracy of classification.

## II. RELATED WORK

Some scholars have proposed APT attack detection methods for DNS traffic. For example, Yan GH et al. [13] analyzed the DNS log, extracted ten features such as host, domain name, and time, and then used unsupervised learning methods to obtain suspicious clusters, and put all domains in the cluster into the list of malicious domain names. The university network data is verified, and all suspicious domain names can be detected. Nadler [14] analyzes the use of tunnels for data transmission in APT attacks, focusing on DNS tunnel detection under low throughput conditions. Some studies [15] have found 7 features related to DNS logs, and proposed a new method to represent the time features of DNS logs, where a neural network has been designed to discover the relationship between DNS activities and APT attacks through deep learning method training model. Some studies [16] observe the DNS traffic between the user's machine and the ISP's local DNS resolver, constructs an annotated bipartite graph, tracks the DNS query behavior of malware infection, and effectively detects the occurrence of new malware control domains. Some studies [17] have proposed a combined method to build an index for DNS traffic and use a predefined filter list for filtering. Then an extensible rule set of abnormal features is defined, and DNS traffic is clustered through the Shared Nearest Neighbor (SNN) algorithm to realize DNS abnormal traffic analysis. Based on the traffic detection results collected from the PANDI CCTLD-ID network, the success rate of using this method to detect abnormal traffic has increased significantly. Some studies [18] put forward an effective method to detect the C&C domain of APT malware with high precision through the analysis of DNS log. First extract 15 features from the DNS log of the mobile device. Score each domain based on Alexa ranking and VirusTotal judgment results. Then select the most normal area based on the scoring criteria. Finally, the Global Anomaly Forest (GAF) algorithm is used to identify the malware C&C domain. Some studies [19] use lightweight open source tools; a combination design of BRO-IDS, LOGSTASH, ELASTICSEARCH and KIBANA Visualization to enhance DNS traffic anomaly detection to obtain real-time performance, better anomaly recognition accuracy and faster detection speed. Some studies [20] have proposed a lightweight botnet detection system based on the Internet of Things (IoT), ConnSpoiler, to detect botnets based on IoT by generating domain streams (AGDs) through rapid identification algorithms. Some studies [21] have established an algorithm for detecting malicious domains in large-scale DNS traffic, namely, Resource Effective Malicious Domain Detector (REMaDD). It does not require prior knowledge of historical malicious activities, uses real-time streaming data from the Inter-University Computing Center (IUCC), and runs on the IBM real-time system. It shows strong performance in terms of detection accuracy and computational efficiency.

Some scholars have proposed detection methods for domain names. Some studies [22] construct a detection method based on the structure of the domain name map, where even if a small-scale seed domain name is used, C&C domain names can be effectively detected. Some studies [23] have proposed a model to judge DGA-DNS detection by analyzing the morphological similarity of ordinary DNS and DGA-DNS, and uses this model to determine the signs of APT attacks and verify its effectiveness. Some studies [24] have proposed a method to detect the two main malicious methods of the Domain Name System (DNS), where a machine learning model is developed to detect information leakage from compromised machines and established a command and control (C&C) server through the tunnel. Some studies [25] have proposed a framework APDD to assist in the detection of APT, using change vector analysis (CVA) and sliding time window methods to analyze the similarity between the access records of the domain name to be detected and the existing APT-related domain names. It then printout a ranking list of suspicious domain name access records, which can be used to manually analyze the most suspicious records first, thereby improving the detection efficiency of APT attacks. Some studies [26] have proposed a graph heuristic algorithm based on belief propagation, which uses the inter-domain relationship in the different stages of APT attack, and infers other attacked hosts and related malicious domain names through known hosts or domain names, so as to achieve early APT Phase detection. The experimental results on LANL simulation attack experiments and a large number of real enterprise Web proxy logs show that this method has higher accuracy and lower false alarm rate. Some studies [27] have proposed a method for detecting malicious domain names based on machine learning. ELM is applied to classify domain names based on features extracted from multiple resources, which can effectively identify malicious domains. Some studies [28] have proposed the MD-Miner system, which takes network traffic and blacklist/whitelist as input to constructs annotated process domain diagrams, to calculate the process behavior features of each domain, and classifies them based on the random forest decision tree model to find new unknown domain. Some studies [29] use natural language processing and machine learning techniques to extract a set of effective features from email clues, users participating in the discussion, and content keywords based on the discussion in the analysis technology mailing list (especially security-related topics), to infer malicious domains from the mailing list. Some studies [30] discover and analyzes the global associations between domains, establishes meaningful associations between domains, and develops a graph-based association domain reasoning technique. Some studies [31] have proposed a DGA domain name detection method based on machine learning. The features of DGA domain names are analyzed through five feature extraction methods, and then six machine learning algorithms and five types of Features sets are applied to obtain 30 candidate DGA detection models. With accuracy, recall rate, F1 score and training time as evaluation indicators, an optimized DGA detection model is obtained through comparative experiments. Some studies [32] have proposed a machine learning-based malicious domain detection system Mal-domDetector to detect DGA-based communications. In addition to using deterministic algorithms, it also uses a set of easy-to-calculate and language-independent features to detect malicious domains.

Some scholars have proposed HTTP feature detection methods. Some studies [33] have proposed an advanced persistent threat detection algorithm (APT) based on HTTP traffic graph model. Some studies [34] have proposed a new C&C feature, namely independent access, to describe the difference between C&C communication and ordinary HTTP requests. The independent access feature is applied to DNS records, a new C&C detection method is implemented, and it is verified on a public dataset. As a new feature of C&C, its advantages and disadvantages are analyzed. Some studies [35] have detected encrypted botnet traffic from normal network traffic and treated it as a traffic classification problem. On the basis of analyzing the features of encrypted botnet traffic, a meta-level classification algorithm based on traffic content features and traffic features is proposed. And wherein the content includes information entropy bytes frequency distribution, traffic features including the port number of the application layer, protocol type, and payload length. It then uses the Naive Bayes classification algorithm to detect botnet traffic. Relevant experiments show that this method has a good detection effect. Some studies [36] have defined behavior features through network traffic patterns. In the proxy server log, select the successful line with the HTTP status code of success, extract the date, time, request line, HTTP status code and size to form a bundle sequence, and extract the feature vector from the bundle. Use Support Vector Machine (SVM) and Random Forest (RF) for classification and prediction. Some studies [37] have proposed a method to distinguish the time transformation features of APT attacks, such as uplink and downlink packet transmission time, downlink average transmission time interval, flow duration, etc., then compare normal traffic and traffic containing malicious loads to capture signals of malicious loads. Some studies [38] have proposed a deep learning (DL) stack to detect advanced persistent threat (APT) attacks. It treats APT as a multi-vector and multi-stage attack with a continuous strategy, using the entire network flow, especially raw data, as the input to the detection process to capture certain types of anomalies and behaviors. Some studies [39] have collected and analyzed network traffic data, and proposed a set of features tailored specifically to detect possible data leakage. A suspiciousness score is defined for each internal host, revealing weak signals related to data leakage and other suspicious APT activities. Some studies [40] have proposed a fractal-based anomaly classification mechanism to monitor and analyze various features of TCP/IP connections, such as the number of packets transmitted, the total number of bytes exchanged, the duration of the TCP/IP connection, and the details of the packet traffic information, effectively reducing false positives and false negatives. Some studies [41] have proposed a framework for detecting real-time traffic of APT attacks, identifying the source of APT attacks based on the collected traffic, and judging whether the attack is a malicious attack through static analysis of the database and dynamic analysis of the sandbox.

Some scholars model multi-type traffic analysis as a detection problem. For example, some studies [42] have proposed an automatic time-dependent traffic detection system (ATCTDS). The evaluation of 36 well-known APT attack datasets shows that the detection of attack behaviors from a series of network attacks involving multiple types has a high accuracy and low false alarm rate. Some studies [43] have proposed a new APT malware detection system based on malicious DNS and traffic analysis. 14 features based on big data are extracted to describe the different attributes of DNS related to malware and its query methods, and network traffic features are defined to identify the traffic of compromised clients that have been remotely controlled. A reputation engine is built to calculate the reputation score of an IP address using these feature vectors together. Some studies [44] have proposed an APT attack detection model based on semi-supervised learning method and complex network features. The entire target network is modeled as a small-world network, and the evolved APT attack network (APT-AN) is modeled as a scale-free network, using a finite state machine to describe the time domain changes of the state during the attack. Some studies [45] have proposed a natural nearest neighbor (less-NNN) method based on fuzzy entropy weighting. First, use Fisher scoring and depth map feature learning algorithm to remove unimportant features and reduce data dimensionality. Then the density of data points, various centers and the minimum radius of the bounding sphere are determined according to the proposed natural nearest neighbor search algorithm (NNN_search). Finally, a fuzzy entropy weight KNN classification method based on closeness is proposed, which improves the accuracy and efficiency of network traffic attack detection. Some studies [46] have proposed a spatio-temporal association analysis method to detect APT attacks in industrial networks, using the FP-Growth algorithm to mine the temporal, spatial and category features of APT attacks, and analyze the relationship between APT attack features. It combines the features of the APT attack history data retrieval method based on the bloomfilter algorithm, explains its laws semantically, and proposes a multi-feature space weighted combination SVM classification detection algorithm to detect abnormal APT attack conversation traffics. Some studies [47] have proposed to use a cascaded LSTMRNNs network to automatically learn features from traffic data for APT multi-step detection. Some studies [48] have proposed an adaptive analysis framework to correlate different types of network security related data, such as network traffic, alarm events and external threat intelligence, and perform real-time detection, event correlation, and attack scenario simplification based on rules and anomaly models. Some studies [49] have proposed a scalable solution Ctracer, which can detect invisible command and control channels in a large amount of traffic data. Some studies [50] have applied machine learning classifiers to network traffic features and proposed an effective multi-level traffic classification method. Some studies [51] have proposed an integrated intrusion detection technology to mitigate malicious events, especially botnet attacks against DNS, HTTP and MQTT protocols used in the Internet of Things. Based on the analysis of the potential features of the protocol, new statistical traffic features are generated. Then, an AdaBoost ensemble learning method is developed using decision tree, Naive Bayes (NB) and artificial neural network three machine learning techniques, to evaluate the effects of these features and effectively detect malicious events.

## III. CONSTRUCTION OF EFFECTIVE SAMPLE SETS

In view of the fact that there are too few effective samples, it is first necessary to construct a sample set for APT traffic detection.

The data source format used in this paper is pcap traffic packet, and the extracted characteristic string includes original fields, statistical behavior, etc. The pcap file format is a commonly used datagram storage format, and mainstream packet capture software including wireshark can generate data packets in this format. DNS and TCP characteristic sequences are generated from the pcap files. For DNS traffic packets, we first call the wireshark API for rule matching to export the required traffic sequence. Then classify and calculate the traffic sequence to generate DNS characteristic sequence. For TCP traffic packets, we use nChronos network performance analysis system to extract features of TCP traffic packets in the unit of session window, and output the session window sequence. Then match a field of features using rules, construct functions to calculate related features, and finally generate TCP feature sequences.

In order to train the proposed model, we construct a data set containing malicious traffic generated by APT samples, which have been determined to belong to the same APT organization. The data set contains DNS traffic and TCP traffic. DNS traffic includes malicious DNS traffic, DNS tunnel traffic and DNS normal traffic.

In addition, this paper also designed another two sets of comparative experimental data to verify the effect of the classifier. The first dataset is mainly used to verify the difference between the analysis of traffic generated by the same APT organization and the analysis of malicious traffic generated by other organizations. On the basis of the original dataset, the traffic generated by other APT tissue samples was added to this group. The second dataset consists of real traffic. It is mainly used to verify whether the APT organization attack can be detected in the actual environment by using the trained model.

When there is an imbalance between positive and negative samples in dataset, the minority class samples are easily overlooked in the classification process. Then classifier is more biased towards the majority class, which leads to the degradation of classification performance. The extracted malicious traffic sequence used in this paper only accounts for 1% to 2% of the total traffic sequence, so it is necessary to use a suitable imbalanced sample processing method to reduce or fill the data. The most commonly used methods include SMOTE[52], Borderline-SMOTE[53], ADASYN[54] and other improved algorithms for traditional oversampling, and EasyEnsemble[55], BalanceCascade, NearMiss and other improved algorithms for traditional undersampling. This paper draws on the EasyEnsemble algorithm to deal with the problem of information loss in undersampling, and improves the ADASYN algorithm for TCP session datasets. The calculation process is shown in Fig. 1.

(1) Calculate sample imbalance as shown in (1):

$$d = \frac{M_{min}}{M_{mid}} + \sum_{type} \frac{M_{type}}{M_{max}} \quad (1)$$

Where $M_{max}$ represents the number of normal samples, $M_{type}$ is the number of abnormal samples, type $\in$ {mid,min}. For DNS abnormal traffic, it includes DNS malicious traffic and DNS tunnel traffic.mid represents the larger number of traffic types in DNS abnormal traffic, and min represents another type. When there are only two types of data in the data set, that is, normal flow and abnormal flow, it degenerates into a two-classification problem. At this time, let min=0. When there is only one data type in the data set, classification is not required at this time. The value of d is set to 0. We set the threshold $d_{th}$ which is the maximum imbalance that can be corrected by the algorithm. When there are two and three types in the data set, the value is different. When $0 < d < d_{th}$ is satisfied, then it will enter (2), otherwise it will exit the program and inform that it cannot be corrected;

(2) Perform multiple sampling with replacement on the normal traffic dataset

This type of dataset accounts for the largest proportion and is a large type of dataset. Each sampling is random sampling. The goal is to generate multiple independent training sets, and then train multiple different generators. Let the number of generators used be n, and the number of white traffic samples randomly drawn by each generator is $\widetilde{M_{max}^{(n)}} = \lfloor \rho * M_{mid} \rfloor$. Denote $\rho$ as the sampling coefficient, then $\rho \in [1,1.5]$, which means that the white traffic must be at least the same as the number of samples in the mid class, and the maximum number selected cannot exceed 1.5 times $M_{mid}$. The calculation steps of each sample generator are as follows, and finally the combined generator $T = \sum T(n)$ is calculated. The calculation of n is based on the number of mid-class samples, as shown in (2):

$$n = \left| \frac{M_{max}}{M_{mid}} \right| \quad (2)$$

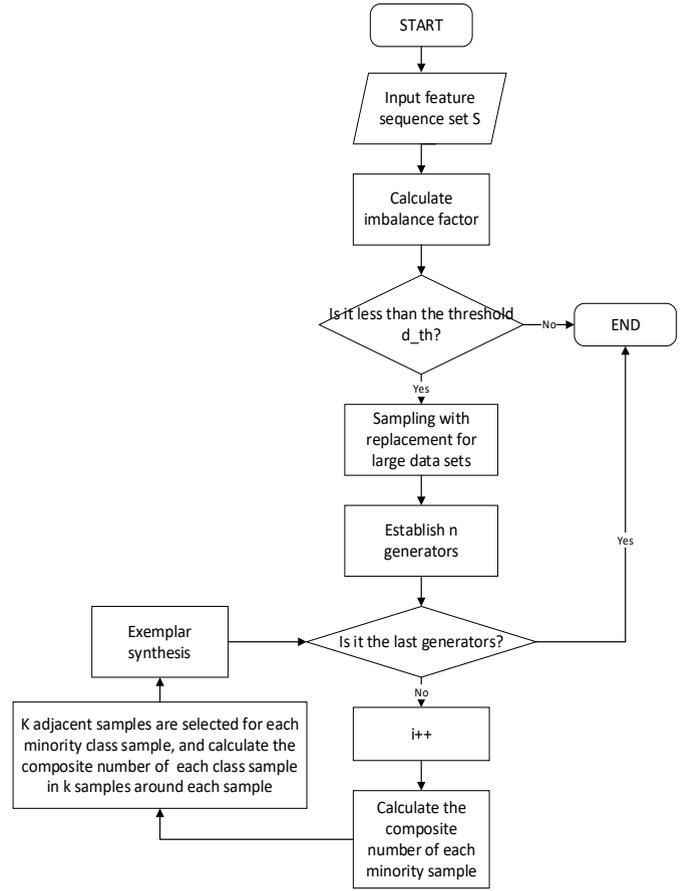

**FIGURE 1.** Flow chart of sample balance calculation.

(3) For each T(n):
Calculate the composite number of each minority sample:

$$G_{type} = \alpha_{type} * (\widetilde{M}_{max}^{(n)} - M_{mid}) \quad (3)$$

$$G_{min\,mid} = (1 - \alpha_{type}) * \widetilde{M}_{max}^{(n)} + \alpha_{type} * M_{mid} - M_{min} \quad (4)$$

Where G represents the number of synthesis, $\alpha_{type} \in (0,1]$, and $\alpha_{mid} = \alpha_{min}$. When $\alpha_{type}=1$, the number of samples synthesized by the two minor classes, $G_{min}$ or $G_{mid}$, is equal to the difference between the normal class and the mid class, and for the mid class the data is just balanced. The min class is then synthesized. $G_{min\_mid}$ is the difference between the number of samples contained in the mid class and the min class, which is also just balanced after adding $G_{min}$ and the original number of min classes. Note that when there are only two types of data, the number of small-type samples is ($\rho$-1)* $M_{mid}$. After determining the number of synthesized samples, go to (b);

(b) Select k nearest samples

In this paper only the minority samples are synthesized. Take each sample $x_{type,i}$ in the abnormal class, and use Euclidean distance to calculate the number of samples belonging to the max class among the k adjacent samples $\Delta_{type,i}$, as well as the number of samples that belong to the non-sample (the minority class contains two categories) around the minority sample $\Delta_{type\_another,i}$. $\Delta_{type\_th}$ and $\Delta_{another\_th}$ are the thresholds for the number of samples in the largest category and the number of samples in other minority categories, respectively. If $k = \Delta_{type,i} + \Delta_{type\_another,i}$ it means that the sample is an isolated sample, and $r_{type,i}$ is not calculated. If $0 \leq \Delta_{type,i} < \Delta_{type\_th}$ or $0 \leq \Delta_{type\_another,i} < \Delta_{another\_th}$, then the number of max class samples or the number of other minority class samples around the sample is lower than the threshold, indicating that the sample is closer to the sample center and the probability of classification error is low. Otherwise, select the samples around the border (for the class two samples, only count the number of samples in the max class in the vicinity of k). This paper believes that the sample imbalance ratio of the border area is close to that of the dataset, and the combined proportion of samples in the border area needs to be greater than the proportion within the class that can be easily classified.

Take the ratio as follows:

$$r_{type,i} = \frac{\Delta_{type,i}}{k} \quad (5)$$

$$r_{\min\_mid,i} = \frac{\Delta_{\min\_mid,i}}{k} \quad (6)$$

The ratio of all samples obtained above is classified and regularized:

$$\tilde{r}_{Type,i} = \frac{r_{Type,i}}{\sum r_{Type,i}} \quad (7)$$

Among them, the Type class means all categories that need to be synthesized.

Calculate the number of samples to be synthesized for each sample in the Type class:

$$g_{Type,i} = \lfloor \tilde{r}_{Type,i} * G_{Type} \rfloor \quad (8)$$

(c) Sample synthesis

Synthesize $g_{Type,i}$ samples for each sample $x_{Type,i}$ in the minority samples.

(d) $for\ j = 1: g_{Type,i}$

Randomly select a Type class sample $\tau_{Type,i}$ from the k proximity of $x_{type,i}$, and calculate the synthesized sample for expansion by $s_j = x_{type,i} + \lambda * (\tau_{Type,i} - x_{type,i})$. Among them, $\tau_{Type,i} - x_{type,i}$ is the distance difference in the multidimensional vector space, $\lambda \in [0,1]$, and each iteration takes any value from the range.

## IV. CONSTRUCTION OF TWO NEW FEATURES

In this section we construct two new features, C2Load_fluct and Bad_rate, to distinguish normal, malicious and DNS traffic of the organization.

### A. RESPONSE PACKET LOAD FLUCTUATIONS C2LOAD_FLUCT

Definition 1: Define the characteristic C2Load_fluct to record the average load of the traffic packet cluster of the same source IP and domain name in the unit domain name within the time window W.

Name the number of DNS response packets of the same source IP in W as k, the mathematical expectation of the packet load from all sample statistics is μ_load, and the domain name length is domain_len, then we have (9) as follows:

$$C2Load\_fluct = \frac{\mu_{load}}{domain\_len} \quad (9)$$

Principle analysis: This feature describes the fluctuation of the average packet load within the time window with the unit length of different domain names. The smaller the calculated value (greater than 1), it proves that the information carried by the domain name per unit length is more relevant to the domain name. It shows that the possibility of using domain names for information transmission is higher. DNS tunnels generally transmit information through TXT requests, and the payload part of the packet body is related to the domain name, so this feature can provide a certain reference for tunnel identification. At the same time, APT attacks generally have the features of periodic visits. In this process, periodic DNS requests for specific hosts and specific domain names will be generated, and periodic response packets will be obtained. Thus, for malicious DNS divided by time window, the calculated characteristic will have certain repeatability.

In order to judge the effectiveness of this feature, it is necessary to judge the correlation between this feature and other features. The two features of Local_abnormal and Response_type contain multiple categories, and one-hot is used for encoding. The heat map obtained after processing is shown in Fig. 2. The horizontal and vertical coordinates in the figure are the features in the feature set, and the intersection indicates the degree of similarity between the two features. The similarity is represented by color, and the corresponding relationship between its value and color is shown on the right

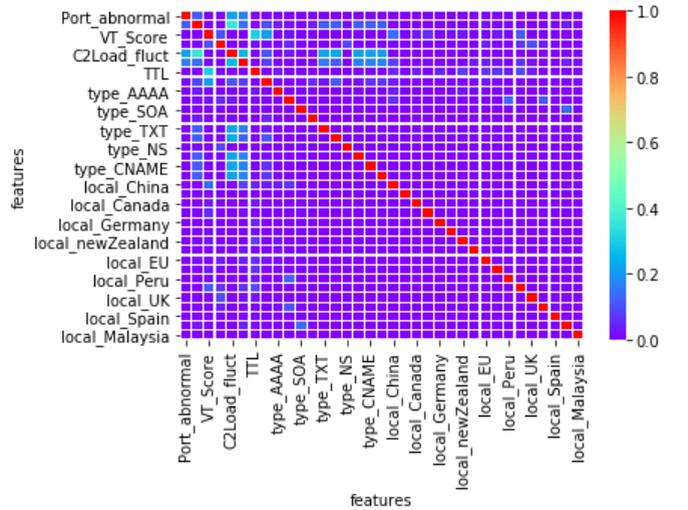

side of the graph.

**FIGURE 2.** DNS feature correlation heat map.

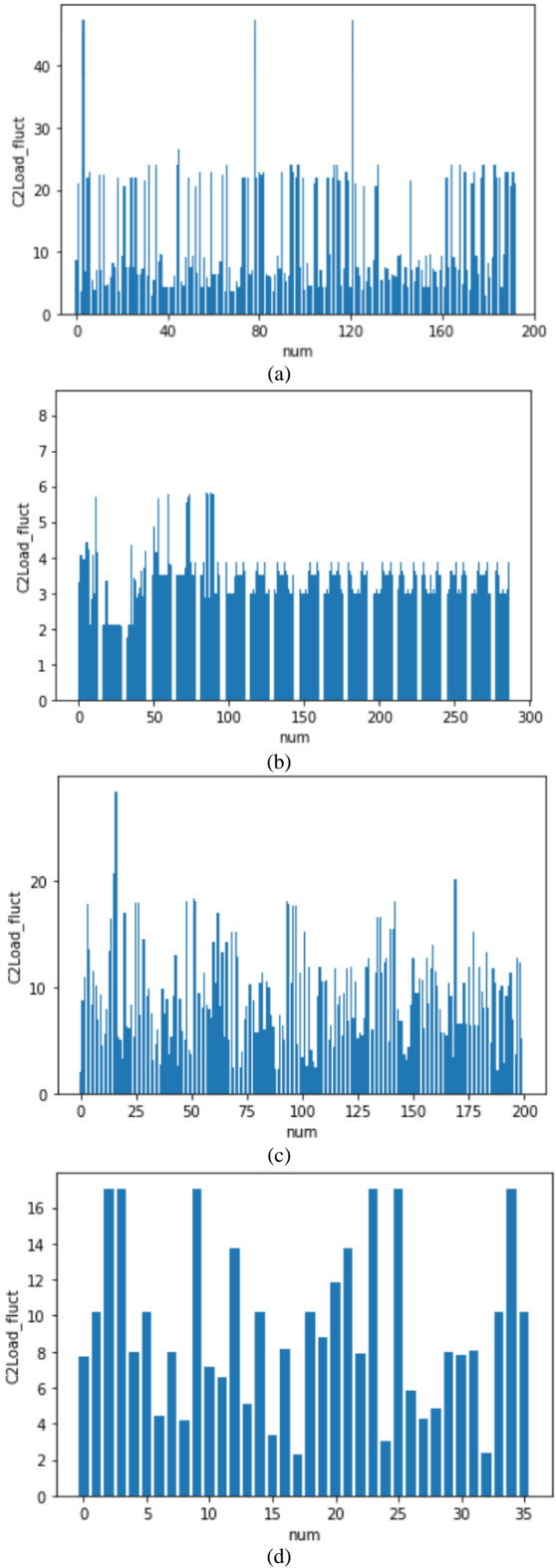

**FIGURE 3.** The distribution of C2Load_fluct in various samples, where the abscissa is the number of DNS feature sequences arranged in chronological order, and the ordinate is the calculated value of C2Load_fluct.

The same feature has the highest correlation. It is expressed as a red point on the diagonal in the figure. As can be seen from the figure, the selected features have extremely low correlation, so they can be used as features for classification. The proposed new features are related to other features The maximum value of the degree is 0.35, which has no strong correlation with other features. So this feature is considered usable.

Fig. 3(a) to 3(d) respectively show the distribution of this feature in various samples, where the abscissa is the number of DNS feature sequences arranged in chronological order, and the ordinate is the calculated value of C2Load_fluct.

It can be seen from Fig. 3(a) that part of the value obtained by malicious traffic statistics is concentrated around 5, the other part is slightly higher than 20, and a small number of calculated values are greater than 40. It can be seen from Fig. 3(b) that the calculated value of DNS tunnel traffic C2Load_fluct generated by the same organization used in this paper is mostly distributed in 1 to 4, and the calculated value of 3 is the most characteristic sequence. For the normal traffic, due to the large amount of data, this paper intercepts the first 200 data of normal traffic and malicious DNS sequences, as shown in Fig. 3(c). By comparison, it is found that the normal traffic rate still has large fluctuations, and no obvious statistical features have been found. For the APT traffic generated by another organization, as shown in Fig. 3(d), it can be seen that there is a certain pattern at the C2Load_fluct values of 17, 06, 10.185, etc., but the data pattern is generally weak. Therefore, the extracted C2Load_fluct is distinguishable for the APT traffic studied in this paper.

### B. BAD_RATE

The new feature Bad_rate constructed in this paper is mainly related to out-of-sequence packets and retransmission packets, and reflects the network status of the APT organization when malicious traffic and normal traffic are generated. The calculation formula is

$$Bad\_rate = \frac{Packet_{out\_order} + Packet_{retransmission}}{Packet_{all}} \quad (10)$$

Where Packet represents the number of data packets in each session window, and the subscript represents whether it is an out-of-sequence packet or a retransmitted packet. Disorder and retransmission are related to the network status at both ends of the traffic. The APT organization uses remote control commands to steal data. In the process of data transmission, it may be affected by network restrictions and C2 end migration, resulting in out-of-order and retransmission of data packets. So the possibility of disorder and retransmission of normal traffic is less than that of APT malicious traffic. Fig. 4(a) to 4(b) show the feature distribution, where the abscissa represents the number of TCP feature sequences arranged in time, and the ordinate represents the calculated Bad_rate value.

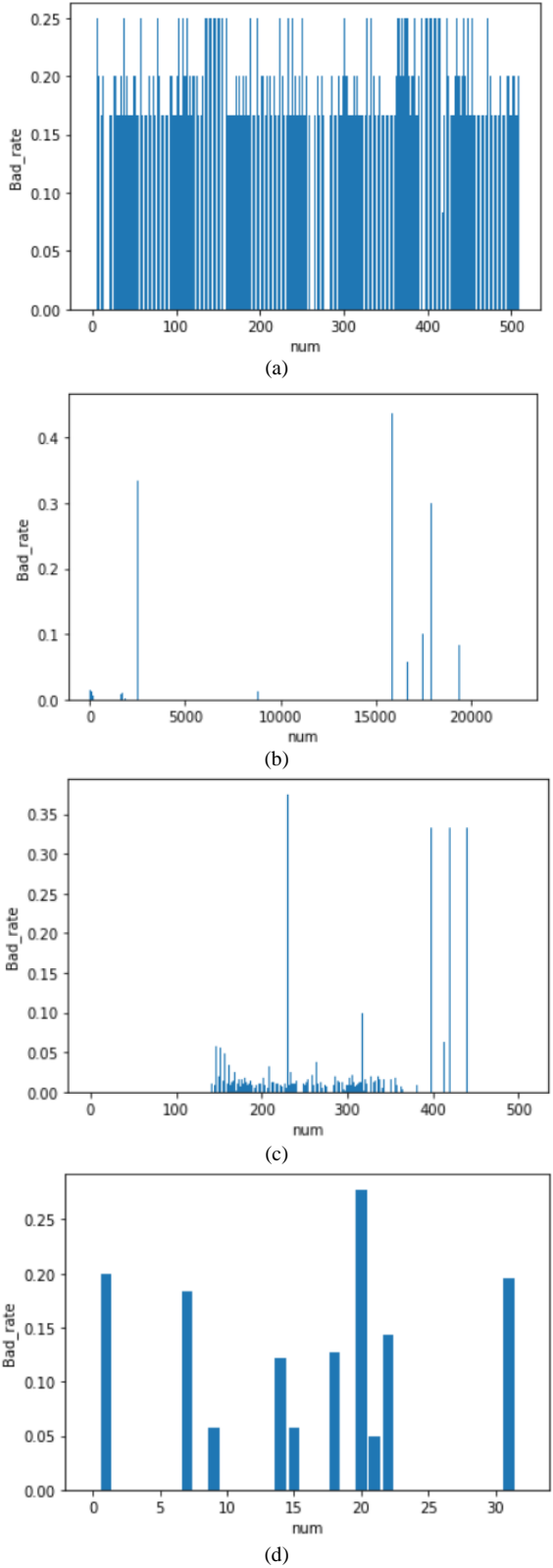

**FIGURE 4.** The distribution of Bad_rate, where the abscissa represents the number of TCP feature sequences arranged in time, and the ordinate represents the calculated Bad_rate value.

Fig. 4(a) and Fig. 4(b) show the TCP Bad_rate distribution of malicious and normal traffic. Analyzing Fig. 4(a) to Fig. 4(b), it can be seen that the bad packet rate of the APT organization traffic is mostly concentrated around 0.17, 0.2, 0.25, and the bad packet rate of normal traffic is more scattered than that of normal traffic. The majority are 0 with a few between 0 and 0.04, and there are some outliers. Because the data volume of normal traffic is too large, in order to strengthen the comparison, Fig. 4(c) intercepts the more densely distributed part of normal TCP traffic, that is, from 1490 to 2000. It can be seen that the dense area is concentrated within 0 to 0.04, which is in line with the previous inference that the value of normal traffic is small. Fig. 4(d) shows the distribution of Bad_rate for other APT organizations. It can be seen from the Fig. 4(d) it's quite a lot of 0 in the bad packet rate of APT organization traffic, so this feature has little effect on the classification of APT organization.

### C. FEATURES SET CONSTRUCTION

In this section the two new features with existing features in existing literature and traffic analysis tools are combined to construct a Features set for APT traffic detection.

#### 1) DNS FEATURES

Table I summarizes the six types of features used to distinguish the malicious DNS traffic of the APT organization from normal traffic. Each type of feature may contain one or more secondary sub-categories, and the new features constructed in this paper are marked with an asterisk (*).

TABLE I
10 DNS FEATURES

| Classification | Abbreviation | Meaning |
|---|---|---|
| Query abnormal | Alexa_score | Alexa website [56] Malicious domain name ranking information |
|  | VT_score | Virus total [57] website test results |
| Port abnormal | Port_abnormal | Port exception |
| Geographic location abnormal | Local_abnormal | The area code of the C2 server |
| Statistics abnormal | C2Load_fluct* | Response packet load fluctuations |
|  | Ask_Res rate | Request response rate |
|  | ClientLen_max | Maximum number of packets sent by the client |
|  | C2Len_max | C2 response packet maximum |
| TTL abnormal | TTL | TTL value |
| Response type abnormal | Response_type | Response packet type |

(1) Alexa_score

The Alexa website counts the comprehensive rankings, visits and other information of websites around the world, which has been recorded as one of the more authoritative website visit evaluation indicators by most people. In the field of APT traffic detection, the number of website visits is often positively correlated with whether the website already exists.

(2) VT_score

Virustotal is a website that provides malware and IOC analysis services. The website integrates multiple detection engines for malware detection websites, such as Symantec, Macfee, 360, etc. Based on this, the returned information from the website can be used to determine whether the IP or domain name is malicious.

(3)Port_abnormal

The main types of Port_abnormal in DNS and TCP are port increment type, port jump type, unconventional port type, etc. In this paper, a weighted linear function is used to express the influence of this value, which is recorded as

$$Port\_abnormal = \sum w_i P_i \quad (11)$$

Where $w_i$ is the weight of various abnormalities in the total abnormality, and $P_i$ is the port abnormality record.

(4)Ask_ Res rate

Malicious DNS and normal DNS have different statistical rules between the number of packets and the relationship between packets in the time window W. Normal user behavior has entropy divergence. The malicious software's DNS packet request has certain regularity due to the attacker's necessity for intranet penetration and data return. If it is a DNS tunnel attack, there will also be certain statistical rules in packet sending and receiving. In the characteristic (12), $num\_Ask_w$ and $num\_Res_w$ respectively represent the number of query packets and the number of response packets for the same source IP to query the same domain name in the time window W.

$$Ask\_Res_{rate} = \frac{num\_Ask_w}{num\_Res_w} \quad (12)$$

(5)TTL

TTL is the retention time of domain name resolution records in the DNS server. According to DNS features used before 2018[58], Leyla et al. [59][60] proposed TTL as a feature for detection. This paper analyzes a dataset of an APT organization and finds that there is also a difference between the TTL distribution of the organization and the normal traffic, so the TTL value is also added as a feature dimension to the classification vector.

(6)Response_type

This paper selects the original record type field in the DNS response packet as a feature. There are many return types for this field, however, this paper discretizes the normal field and mainly includes ten types of records such as A, AAAA, CNAME, MX, TXT, etc. This feature is selected to distinguish the traffic belonging to the DNS tunnel from the malicious traffic. For example, if an attacker uses dnscat2 and its equivalent software for tunnel transmission, CNAME, MX, and TXT records may be generated. Generally speaking, DNS tunnels are more inclined to use TXT records for data return, because TXT records can return more data than normal records.

2) TCP AND HTTP/HTTPS FEATURES

The features of TCP and HTTP/HTTPS selected in this paper are shown in Table II. Among them, Bad_rate is a new feature constructed in this paper, ∘ marks the feature after the second summary, and ∆ marks the features other than the features given by the software.

Note: There are five types of TCP connection status: established, closed, syn_sent, time_wait, and syn_received.

(1)Duration_T

Statistical analysis found that the TCP session window duration of the APT is different from that of normal traffic. As shown in Fig. 5(a) and Fig. 5(b), the abscissa of the picture represents the duration (in seconds), and the ordinate represents the number of conversation windows. The analysis found that most of the session windows of normal traffic lasted within 0-5s, and showed a rapid downward trend. The number of session windows in 0s was the largest, exceeding 20,000. There are fewer session windows lasting more than 5s, and the longest duration is 40s. Correspondingly, the malicious traffic curve is non-smooth. Most samples last within 0-6s. Unlike normal traffic, most of the session windows are at 1s, and only 36 are at 0s. At the three odd-numbered second time points of 1s, 3s, and 5s, the number of malicious traffic session windows reached a local peak.

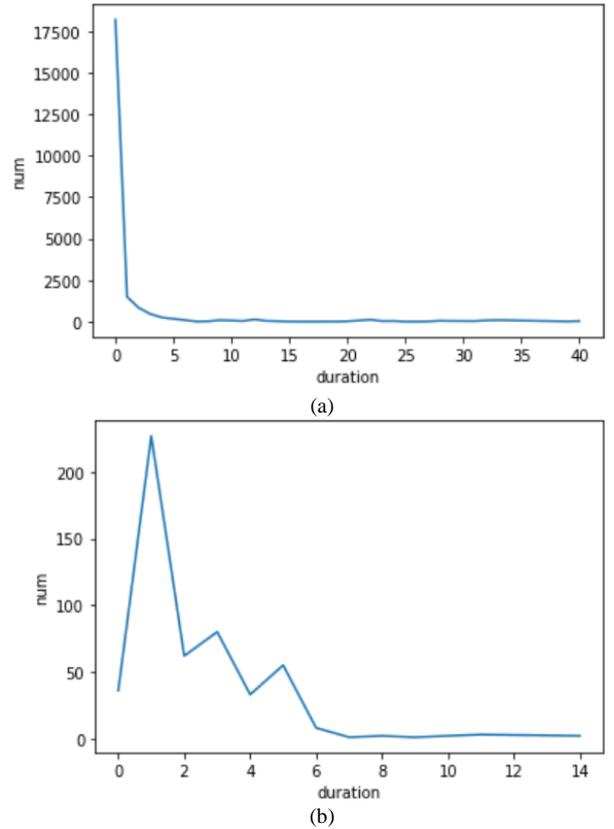

**FIGURE 5.** Duration of normal TCP traffic and duration of malicious traffic from the APT organization.

TABLE II
FEATURES OF 11 TCP SESSION SEQUENCE

| Classification | Abbreviation | Meaning |
| --- | --- | --- |
| Query abnormal | Alexa_score$^\Delta$ | Alexa ranking |
| | VT_score$^\Delta$ | Virustotal website test results |
| Geographic location abnormal | Local_abnormal | The area of the C2 server |
| Time abnormal | Duration_T | TCP session window duration |
| Statistics abnormal | Bad_rate° | Bad packet rate |
| | Upload_num | Number of upload packets |
| | Upload_load | Upload packet load |
| | Upload_numRate° | Upload packet rate |
| | Upload_loadRate° | Upload load rate |
| Port abnormal | Port_abnormal° | Port abnormal |
| TCP abnormal | TCP_connectState | TCP connection status |

(2)Upload_num and Upload_load

These two values refer to the original statistical values given by the Network Performance Analysis System [61], which can be used to determine the difference between the number of uploaded packets and the size of the uploaded packets in the characteristic sequence of each session window of the normal traffic and the APT traffic of the organization.

(3)Upload_numRat and Upload_loadRate

Compared with the use of the four original features of the number of uplink data packets, the number of downlink data packets, the uplink packet load, and the downlink packet load, the Upload_numRate and Upload_loadRate features record the statistical law between the uplink data packet and the downlink data packet in a single session window. Upload_numRate is the ratio of the number of uplink data packets to the total number of packets in a single session window. Compared with the normal traffic of APT traffic, the number of uplink data packets is greater than the number of downlink data packets. Analysis of the APT life cycle shows that when C2 is connected, it needs to get instructions from the remote to connect, and when the life cycle ends, it needs to send instructions to self-destruct. In addition, most of the host side connects to the C2 side, including many data packets that cannot be transmitted to the target host and data leakage. Upload_loadRate is the ratio of the uplink load to the total load. Similarly, the purpose of most APT activities is data theft, so the average load value of uplink data packets in the same session window is much larger than that of downlink packets.

(4)TCP_connectState

This feature records the stage of the TCP connection in the selected session. As a feature, it is used for TCP traffic classification here. Some studies[62] pointed out that TCP connection is divided into 5 stages. Analysis of the data of normal traffic and malicious traffic shows that the stage of normal traffic is mostly closed. APT malicious traffic has a lower rate of successful C2 connection, and there are relatively more sessions in the syn_send and syn_received stages compared with normal traffic.

## D. TRAFFIC CLASSIFICATION BASED ON FEATURES SET

In order to be applicable for small sample APT traffic detection, this paper optimizes the Adaboost algorithm.

The AdaBoost algorithm is essentially an iterative process in which a base learner is established from unweighted samples to generate class labels. When the problem is multi-classification, the accuracy of random guessing is only 1/K (for example, the accuracy of the three classifications of DNS traffic characteristic sequence in this paper is 1/3 in the case of random guessing), and it may be much more difficult to guarantee an accuracy of 0.5 than 1/K. Therefore, if the weak classifier is not properly selected, the failure of AdaBoost will increase accordingly.

Referring to the SAMME algorithm proposed by Ji Zhou[63] in 2009, this paper changes the weight coefficient of the base learner update from $\log \frac{1-\varepsilon_t}{\varepsilon_t}$ to $\log \frac{1-\varepsilon_t}{\varepsilon_t} + \log(K - 1)$. In this way, as long as $1 - \varepsilon_t > 1/K$ is guaranteed, the trained classifier can be made greater than the accuracy of random guessing instead of 0.5, which is in line with the original AdaBoost idea of setting the coefficients in the two classifications. In summary, the AdaBoost three classification process used in this paper is as follows:

(1)Let S be the total number of sample training sets, X is the sample space, and Y is the category space. Then we have:

$$S = \{(x_i, y_i) | i = 1,2, \dots, m, x_i \in X, y_i \in Y\} \quad (13)$$

(2)Initialize the weight of the training sample, the weight of the first round of sample x_i is:

$$w_{1i} = \frac{1}{m} \quad (14)$$

The weight space it belongs to is $D_1 = (w_{11}, \dots w_{1i}, \dots w_{1m})$

Perform T rounds of iteration: t=1,2,...T. Each iteration uses the base learner $h_t$, and the error rate of the base learner on its training set is as follows, where $I(h_t(x_i) \neq y_i)$ represents the situation where the base learner $h_t$ classifies the sample $x_i$ incorrectly:

$$\varepsilon_t = P(h_t(x) \neq y) = \frac{\sum_i^m w_{ti} I(h_t(x_i) \neq y_i)}{\sum_i^m w_{ti}} \quad (15)$$

The weight coefficient of the base learner $h_t$ is denoted as $\alpha_t$, and the modified $\alpha_t$ calculation formula is:

$$\alpha_t = \ln \frac{1 - \varepsilon_t}{\varepsilon_t} + \ln(K - 1) \quad (16)$$

New sample distribution $D_{t+1} = (w_{t+1,1}, \dots w_{t+1,i}, \dots w_{t+1,m})$, where the calculation formula of $w_{t+1,i}$ is as follows:

$$w_{t+1,i} = w_{t,i} e^{\alpha_t I(h_t(x_i) \neq y_i)} \quad (17)$$

Normalize $w_{t+1,i}$:

$$w_{t+1,i} = \frac{w_{t+1,i}}{\sum_i^m w_{t+1,i}} \quad (18)$$

Get the final integrated classifier:

$$H(x) = sign\left(\sum_{t=1}^T \alpha_t I(h_t(x_i) = y_i)\right) \quad (19)$$

## V. EXPERIMENTS AND RESULTS ANALYSIS

The experimental environment used in this paper is a personal PC, the PC operating system is macOS, version 10.14.3, the processor is 2.3 GHz Intel Core i5, and the memory is 16GB. The development tool is Anaconda3, and the Python version is 3.6.

### A. EXPERIMENTAL PLATFORM ARCHITECTURE

The overall module structure of the APT testing experimental platform is shown in Fig. 6.

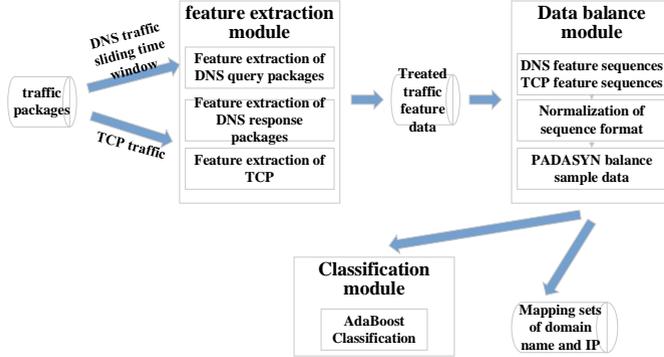

**FIGURE 6.** Overall module structure diagram

The input traffic packet is the processed DNS, TCP, http / HTTPS traffic. Then the data formatting module is imported in two directions. The DNS traffic is sent to the DNS data format sorting module. This function classifies the query packet and the response packet through the QR field of the packet header and extracts the statistical features of the two types of packets on the time window to generate the characteristic sequence. Use network performance analysis system to filter TCP and HTTP/HTTPS traffic into the characteristic sequence of TCP session windows. The above two types of sequences are sorted and stored in the traffic characteristic database. The data preprocessing module extracts data from the traffic feature database, performs secondary statistical feature extraction, generates the corresponding feature sequence, and balances the feature sequence set of the sample through the PADASYN algorithm. The generated feature sequence is input to the 3-class classifier. The data preprocessing module also establishes a domain name IP mapping set by extracting the domain name IP relationship in the traffic.

### B. DATASETS

This paper designs three datasets to test our proposed method. Among them, dataset 1 is composed of white traffic generated by normal software and malicious traffic generated by APT samples. Malicious traffic is generated by multiple APT samples in an isolated environment composed of sandboxes and real hosts. These samples have been marked by manual dynamic and static analysis and have been determined to belong to the same APT organization. White traffic is the background traffic generated by other normal software in the same period. In order to verify the effect of unique characteristics on different APT organizations, we added the traffic generated by other APT tissue samples on the basis of dataset 1, forming dataset 2. Dataset 3 is the real traffic from Ansai APT advanced threat analysis platform. However, the traffic generated by the gateway connected to the platform is huge, and all the data cannot be marked. Therefore, this dataset is used as a comparative test set to verify whether the method in this paper can detect the APT organization attack in the actual environment.

We divide the dataset 1 into the training set and test set according to the ratio of 8:2 for training and testing our model. During the experiment, after importing the traffic in dataset 1 into the data formatting module, the original feature sequence is generated. The included DNS original feature sequences totaled 81,514, and TCP session window sequences totaled 22,828. Further process the feature sequence, extract the required features, and reduce the dimensionality of the original DNS sequence. The final DNS feature sequence generated by statistics is 31632, which is 38.81% of the original DNS feature sequence. The ratio of malicious DNS traffic, DNS tunnel traffic and DNS normal traffic is 2:3:310.

### C. RESULTS ANALYSIS

This paper uses the AdaBoost classifier and compares the classifiers to choose the decision tree, KNN, Naive Bayes and SVM algorithm proposed by Chu WL [64] and Ghafir Ibrahim [65]. At the same time, due to the imbalance of positive and negative samples in the dataset, this paper uses F1-score as an evaluation index to evaluate the classification effect. For dataset 3, due to the excessive number of feature sequences in the dataset, it cannot be fully verified manually. For this reason, this paper only verifies the feature sequences classified as abnormal. In addition, since it is unclear whether the characteristic sequence classified as normal contains malicious sequences, it can only determine how many normal sequences are incorrectly marked and how many malicious sequences are correctly marked in the characteristic sequences classified as abnormal. Therefore, for actual traffic, this paper uses Precision (precision rate, the probability of a malicious sequence in a sequence predicted to be malicious) as an evaluation indicator.

For the DNS traffic sequence, the results when using and not using the C2Load_fluct feature are shown in Table III under dataset 1 and dataset 2 and five classifiers.

Table III shows that for the DNS traffic in dataset 1, the classification effect obtained by using C2Load_fluct in the above classifier is better, where it is improved by around 2% on SVM (RBF), Naive Bayes, and AdaBoost. The classification effect of other algorithms is also slightly improved. In addition, during the experiment, the SVM-linear calculation speed is very slow (here, DNS is 3-class classification), so in practical applications, the use of this classifier for classification should be avoided as much as possible. For dataset 2, it can be seen that the detection effect can also be improved after using C2Load_fluct, but the overall result is poor compared to dataset 1.

TABLE III
COMPARISON OF F1-SCORE WITH C2LOAD_FLUCT AND WITHOUT C2LOAD_FLUCT

| | Grouping\Classification Algorithm | SVM (linear) | SVM (RBF) | Naive Bayes | Decision Tree | KNN | AdBoost |
|---|---|---|---|---|---|---|---|
| 1 | With | 0.7500 | 0.6512 | 0.9470 | 0.9621 | 0.9589 | **0.9827** |
| | Without | 0.7411 | 0.6294 | 0.9287 | 0.9589 | 0.9553 | 0.9673 |
| 2 | With | 0.7007 | 0.6364 | 0.9101 | 0.9320 | 0.9333 | **0.9419** |
| | Without | 0.6932 | 0.6271 | 0.8826 | 0.9249 | 0.9238 | 0.9355 |

For the TCP session feature sequence, Table IV shows the classification results of the five classifiers in the dataset 1 and dataset 2 when using nChronos network performance analysis system and the 11 TCP features in this paper (including the new Bad_rate feature). It can be found that compared to using the nChronos network performance analysis system directly, selecting the feature group in this paper has a better effect on the APT organization session window classification.

TABLE IV
COMPARISON OF DIFFERENT FEATURES OF TCP SESSION SEQUENCE UNDER DIFFERENT DATASETS

| | Grouping\Classification Algorithm | SVM (linear) | SVM (RBF) | Naive Bayes | Decision Tree | KNN | AdBoost |
|---|---|---|---|---|---|---|---|
| 1 | nChronos | 0.7235 | 0.7823 | 0.8944 | 0.9309 | 0.9238 | 0.9561 |
| | In this paper | 0.7809 | 0.7982 | 0.9115 | 0.9694 | 0.9336 | **0.9898** |
| 2 | nChronos | 0.7326 | 0.7508 | 0.8505 | 0.8920 | 0.8686 | 0.9272 |
| | In this paper | 0.7526 | 0.7730 | 0.8601 | 0.9251 | 0.8886 | **0.9469** |

Next, we use the model trained in dataset 1 to classify dataset 3. The results are shown in the Table V. It can be found that compared to using the nChronos network performance analysis system directly, selecting the feature group in this paper has a better effect on the APT organization session window classification.

TABLE V
PRECISION COMPARISON OF DIFFERENT FEATURES USED IN ACTUAL TRAFFIC

| | Grouping\Classification Algorithm | SVM (linear) | SVM (RBF) | Naive Bayes | Decision Tree | KNN | AdBoost |
|---|---|---|---|---|---|---|---|
| DNS | With | 0.8183 | 0.7470 | 0.7226 | 0.8105 | 0.7304 | **0.8426** |
| | Without | 0.8123 | 0.7414 | 0.7022 | 0.8047 | 0.7283 | 0.8314 |
| TCP | nChronos | 0.7531 | 0.8424 | 0.7125 | 0.7966 | 0.7851 | 0.8317 |
| | In this paper | 0.8093 | 0.8538 | 0.7237 | 0.8152 | 0.8019 | **0.8645** |

Since the classifier is an AdaBoost decision tree, n_estimators=200 is fixed in this paper to debug learning_rate. The DNS and TCP data on the two datasets are adjusted separately, as shown in Table VI.

The analysis of Table VI shows that for dataset 1, the F1-score calculated for the DNS sequence and TCP sequence is the largest when learning_rate=1.0 and learning_rate=0.8, respectively. For dataset 2 with the traffic of other organizations added, the F1-score calculated for its DNS sequence reached 0.9419 when learning_rate=0.4.

TABLE VI
LEARNING_RATE PARAMETER ADJUSTMENT

| | | F1-score | | | |
|---|---|---|---|---|---|
| | | Dataset 1 | | Dataset 2 | |
| Frequency | learning_rate | DNS | TCP | DNS | TCP |
| 1 | 0.2 | 0.9597 | 0.9458 | 0.9031 | 0.9230 |
| 2 | 0.4 | 0.9597 | 0.9645 | 0.9419 | 0.9469 |
| 3 | 0.6 | 0.9598 | 0.9722 | 0.9223 | 0.9197 |
| 4 | 0.8 | 0.9442 | 0.9898 | 0.9223 | 0.9069 |
| 5 | 1.0 | 0.9827 | 0.9898 | 0.8586 | 0.9069 |

In addition, dataset 2 is weaker than dataset 1 in the overall classification effect. Dataset 2 is constructed by adding a small amount of malicious traffic generated by other APT organizations on the basis of dataset 1. Through the analysis of the feature extraction results in 5.2.1, we can see that there are features that are not sensitive to the APT tissue traffic in the features used in this paper. At the same time, because the feature dimension used in this paper is relatively low, there may be features that are sensitive to the traffic of the APT organization outside the feature space divided herewith. Therefore, when using the classification model to classify APT attack traffic, it is best to analyze the traffic generated by the same organization and extract a combination of features that can represent the commonality of the organization's traffic.

Taking dataset 1 as an example, the data distribution after the APT organization traffic classification is shown in Fig. 7 and Fig. 8. In Fig. 7, red, yellow, and blue respectively represent normal, malicious, and DNS tunnel traffic. The 2D diagram shows the selected abscissa is Ask&&Res rate, and the ordinate is TTL(all features are normalized). As shown in Fig. 7, the statistical law between ask and request packets of normal traffic is weak. But DNS tunnels and malicious DNS have local clusters for this feature. In Fig. 8, red and blue represent TCP normal and malicious traffic respectively. The 2D diagram shows that the selected abscissa feature is Upload_numRate and the ordinate is Bad_rate. It can be seen from Fig. 7-8 that the upload packet rate of normal traffic is relatively small, and the upload packet rate of malicious traffic is relatively high. At the same time, the bad packet rate of normal traffic is mostly within 0.05 (corresponding to the position of the coordinate axis 0.2), and the bad packet rate of malicious traffic is mostly distributed around 0, 0.17, 0.2, and 0.25.

For dataset 3, due to data labeling issues, only the Precision can be analyzed. It can be seen from Table 5 that after classification using the model trained with simulated traffic, its accuracy rates are 0.8426 and 0.8645 respectively in the DNS and TCP datasets. It shows that under the two datasets marked as malicious, there are some misclassified white traffic. It is guessed that because the actual dataset is larger, it contains more white traffic, and because it is impossible to obtain statistical rules for the white traffic, some of the traffic may have similarities under the features of this paper. Therefore, in order to strengthen the classification effect, future research needs to expand both the data source and feature dimensions.

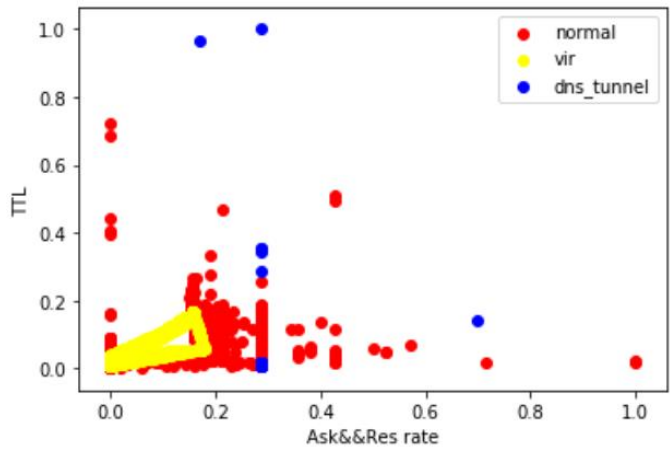

**FIGURE 7.** DNS classification result.

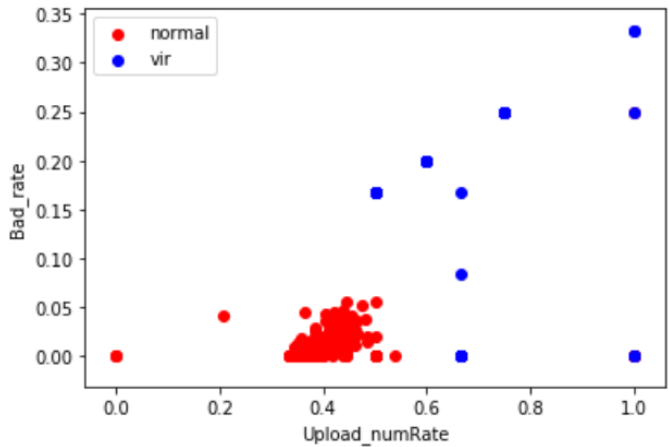

**FIGURE 8.** TCP classification result.

## VI. FURURE WOEK

Due to the impact of environmental resources, this paper does not analyze the massive real-time data. In addition, it only considers the traffic, and does not involve the research of APT malicious code, host behavior, etc. Therefore, we plan to advance the following work:

(1) Map the traffic and attack phases through the kill chain model and causal scenario generation algorithm. The generation algorithm connects the traffic nodes and states of the same IP and domain names, so that security personnel can make a more reasonable assessment of the APT attack process.

(2) Combine the traffic detection method proposed in this paper with previous studies on malicious code and host behavior to conduct a larger-scale correlation analysis. This will be able to extend the kill chain to more stages and improve the detection efficiency of the organization's APT attacks.

## VII. CONCLUSION

In the feature selection, window detection is used in this paper to extract statistical features for DNS traffic, distinguishes DNS tunnel traffic in the organization's APT traffic, and divides DNS tunnel traffic into the data leakage stage. Whether the extracted TCP feature sequence contains HTTP/HTTPS traffic is determined based on TCP stage in the feature. New features with the features of the APT organization are constructed for both types of traffic, which are response packet load fluctuation (C2Load_fluct) and bad packet rate (Bad_rate). Three types of traffic are classified for a certain APT attack on the basis of constructing a Features set containing two new features, where the experimental results show that the F1-score value of the classification can reach above 0.94.

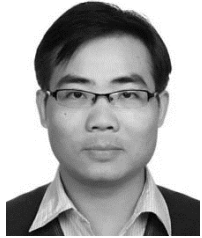

**LEI GUAN** PhD Candidate in School of Electrical Engineering, Tsinghua University, and received the master's degree from school of computer science, Beijing University of Technology, China, in 2010. His research interests cover IoT security technology, social application network analysis.

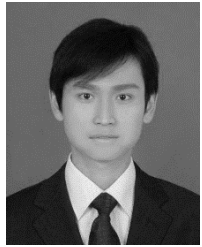

**HAN GUO** Engineer State Grid Information & Telecommunication Branch, and received the master's degree from University of Chinese Academy of Sciences, in 2018. His research interests cover IoT security technology, computer technology.

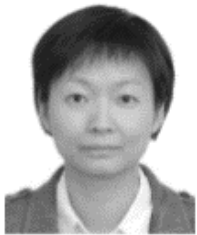

**RU ZHANG** professor at Beijing University of Posts and Telecommunications. She received the Ph.D.degrees from school of comuputer science, Beijing Institute of Technology in 2003. Her research interests cover multimedia security, watermarking, and vulnerability analysis.

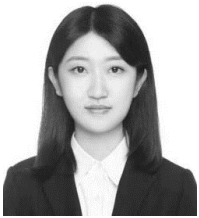

**WENXIN SUN** received the M.S. degree from Beijing University of Posts and Telecommunications in 2020. Her research interests include watermarking, deep learning and Information security.

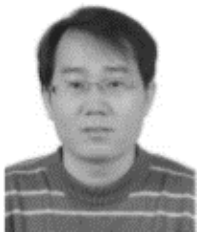

**JIANYI LIU** received the B.S. degree from Xian University of Posts and Telecommunications in 2000 and Ph.D in engineering from Beijing University of Posts and Telecommunications in 2005. His current research interests cover information security, natural language processing, and big data analysis, etc.

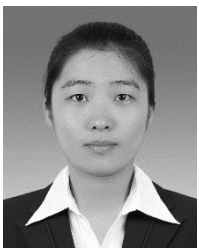

**JINGWEN LI** received the B.S. degree from Harbin Engineering University in 2019. She is currently working toward the Ph.D. degree at the Beijing University of Posts and Telecommunications. Her research interests include information security, natural language processing, and big data analysis.